\documentstyle[aps,epsf]{revtex}

\begin{document}

%\wideabs{

\title{Dynamics of Elastic Boundaries}

\author{Sharad Ramanathan and Alexander E.~Lobkovsky}
\address{Institute for Theoretical Physics, University of California,
  Santa Barbara, CA 93106}

\date{\today}

\maketitle

\begin{abstract} 
  We study, both analytically and numerically, the dynamics of elastic
  boundaries such as crack fronts in fracture and surfaces of contact
  in solid on solid friction.  The elastic waves in the solid give
  rise to kinks that move with a characteristic velocity along the
  boundary.  As stopping kinks pass through they cause moving parts of
  the boundary to stop.  Starting kinks cause stationary parts of the
  boundary to move.  We study the interaction of these kinks with
  disorder that arises from the spatial variations of the friction
  constant or fracture energy.  In the absence of elastic waves,
  elastic boundaries with disorder operate at a critical point leading
  to a power-law distribution of slip events and self-affine
  boundaries.  Elastic waves result in relevant perturbations at this
  fixed point.  Slip events beyond a critical size run away and the
  velocity of the boundary jumps to a nonzero value when the external
  load is increased above a threshold.  We analyze in detail a class
  of simple models that capture the essential features of bulk
  vectorial elasticity and discuss the implications of our results for
  friction and fracture dynamics.
\end{abstract} 
%}

Moving elastic boundaries occur in a wide range of problems.  Examples
are the interface of contact in solid on solid friction, the crack
front in fracture, or the line of contact in the case of peeling of an
elastic film off a substrate.  There remain several unsolved puzzles.
For instance, why the frequency of the earthquakes averaged over all
faults is proportional to a power of the seismic moment \cite{guten},
while some faults show an excess of large events.  Power law
statistics of slip events may imply that the boundary is rough on all
length scales just before it begins to move.  Experiments on cracks
restricted to move in a plane between two pressure welded Plexiglas
sheets find that the crack front is self affine near onset of fracture
\cite{schmittbuhl}.  The dynamics of an earthquake fault during a
single slip event has received much attention since the inversion of
seismic data by Heaton \cite{heaton} discovered slip bands of finite
spatial extent.  Nielsen \textit{et al} \cite{stefan} found
corroborating numerical evidence in simulations of faults with rate
and state dependent friction.  Both studies use acoustic models,
neglecting the vectorial elastodynamics of the bulk.  Details of how
these slip bands arise and how earthquakes start and stop are yet
poorly understood.

Analytic progress in problems with moving elastic boundaries is
difficult and numerical simulations are intensive.  Fisher and
collaborators \cite{dsf} considered a class of models to study rupture
along heterogeneous earthquake faults.  These were found to naturally
operate at a critical point leading to power-law scaling of earthquake
statistics. They also considered dynamical effects arising due to
friction laws and elastic waves in an acoustic model.  A variety of
other simplified models to study the dynamics of boundaries omit some
important features such as long range elasticity \cite{bk}, bulk
elastic waves \cite{quasistatic} and the vectorial nature of the
problem \cite{dsf,myers}.  To reproduce the complexity in the dynamics
of friction for instance, the transition between stick slip and steady
sliding, Carlson and Batista introduced complicated friction laws
\cite{rateandstate}.

In this letter, we show that vectorial elastodynamics of the bulk
results in complex friction dynamics even when the boundary obeys
Coulomb friction.  We find kink solutions that move along the boundary
with a specific velocity and identify generic features of the
equations of motion for elastic boundaries that would support such
solutions.  To understand the interactions of kinks with
heterogeneities, we study a class of models, in the spirit of
Re f.~\cite{dsf}, which capture the effects of elastodynamics.  In our
models, these kinks qualitatively alter the dynamics of the boundary
near the onset of motion.

\begin{figure}[htbp]
  \begin{center}
    \vskip 0.5in
    \centerline{\epsfxsize=5in \epsfbox{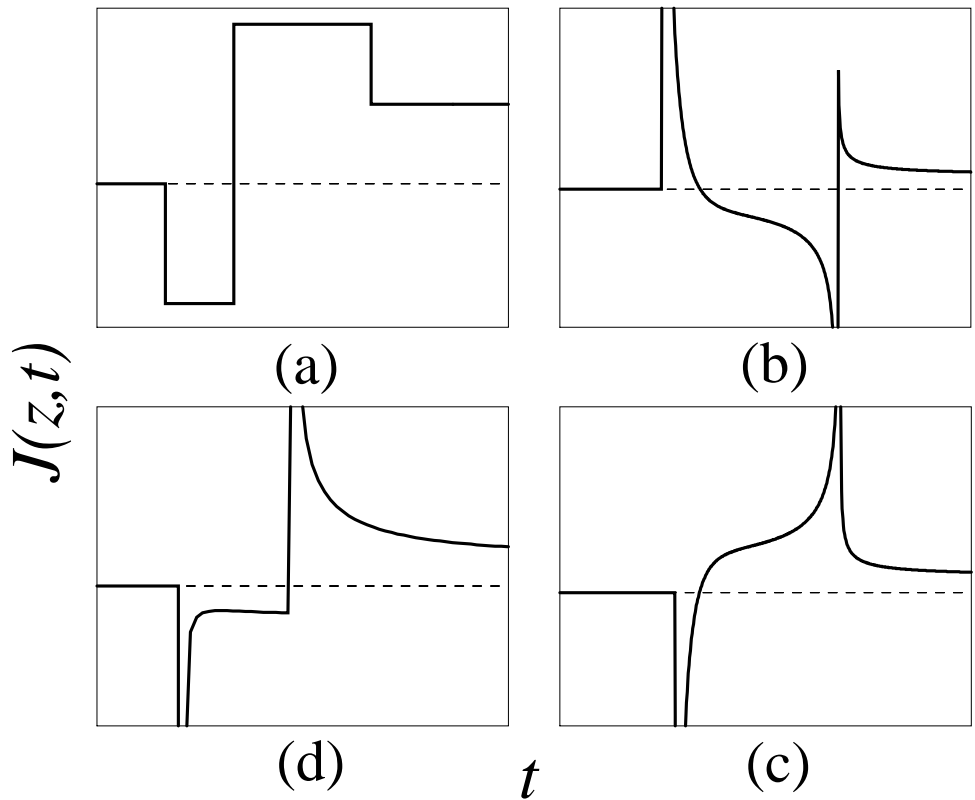}}
    \caption{Kernel $J(z, t)$ at position $z$ as a function of time
      $t$.  Model kernel (a) represents essential features of
      kernels that arise in fracture (b) and solid on solid friction
      (c) and (d).  Graph (c) is for $z < 0$, (d) is for $z > 0$.  The
      friction coefficient is $\mu = 3$.}
    \label{fig:kernels}
  \end{center}
\end{figure}

To motivate our study, let us consider the problem of an elastic solid
sliding on and always in contact with a flat infinitely rigid
substrate.  Let $\hat y$ be normal to the substrate and $\hat z$ be
the direction of sliding.  We study a two dimensional problem by
requiring translational invariance along the $\hat x$ direction.  Let
$f(z, t)$ be the displacement in the $z$ direction of the solid's
boundary relative to the substrate.  An external shear stress
$\sigma^\infty_{\rm yz} = T$ and a normal load $\sigma^\infty_{\rm yy}
= P$ are applied far away.  We obtain an effective equation of motion
for $f(z, t)$ by requiring that on the boundary,
\begin{equation}
  \label{eq:fricgen}
  \cases{
  \sigma_{\rm yz} = \mu \sigma_{\rm yy}, & when the solid is sliding,
  $\dot f > 0,$ \cr
  \sigma_{\rm yz} < \mu \sigma_{\rm yy}, & when it is stuck, $\dot f =
  0.$}
\end{equation}
(Note that the solid can slide only in the positive $z$ direction,
hence $\dot f \ge 0$).  We assume that the friction coefficient $\mu$
is independent of slip or sliding velocity \cite{other_laws}.  Given
$f(z, t)$ and that elastic waves that radiate into the bulk are not
reflected back, one can solve the equations of elastodynamics to
obtain the resulting stresses at the boundary.  Then,
Eq.~(\ref{eq:fricgen}) becomes
\begin{mathletters}
  \label{eq:general}
  \begin{equation}
    {1 \over 2}{\partial f(z, t) \over \partial t} = \Sigma(z,
    t)\,\Theta(\Sigma(z, t)),
  \end{equation}
  where,
  \begin{equation}
    \label{eq:stress}
    \Sigma(z, t) = \int_{z', t' < t} dz'dt' \, J(z - z', t - t') \,
    {\partial f(z', t') \over \partial t'} + T - \gamma,
  \end{equation}
\end{mathletters}
with $\gamma = \mu P,$ and $J = J_{\rm friction} = J_{\rm yz} - \mu
J_{\rm yy},$ where
\begin{equation}
  \label{eq:Jyz}
  J_{\rm yz} = \delta'(z) + {1 \over 2\pi} {\partial
    \over \partial t} \cases{ \displaystyle{{1 \over t} {\sqrt{1 -
          z^2/\kappa t^2} \over 1 - \sqrt{1 - z^2/t^2}\sqrt{1 -
          z^2/\kappa t^2}}}, & for $0 \le |z|/t < 1$ \cr
    \displaystyle{{\sqrt{1 - z^2/\kappa t^2} \over t \left[1 +
          (z^2/t^2 - 1)(1 - z^2/\kappa t^2)\right]}}, & for $1 < |z|/t
    < \sqrt{\kappa},$\cr}
\end{equation}
and
\begin{equation}
  \label{eq:Jyy}
  J_{\rm yy} = {1 \over 2\pi t} {\partial \over \partial z} \left[{z/t
      \sqrt{z^2/t^2 - 1} \sqrt{1 - z^2/\kappa t^2} \over 1 + (z^2/t^2 -
      1) (1 - z^2/\kappa t^2)}\right], \qquad {\rm for} \quad 1 < |z|/t
  < \sqrt{\kappa}.
\end{equation}
Here $\kappa$ is the square of the ratio of the longitudinal to the
transverse bulk wave speeds.  A plot of the friction kernel $J_{\rm
  friction}$ as a function of time $t$ at a fixed $z$ is shown in
Figs.~\ref{fig:kernels} (c) and (d).

When $T>\gamma$, there exists a unique steady \cite{footnote2} state
with the whole boundary moving at a constant velocity.  We now seek,
for $T<\gamma$, steady state solutions of Eq.~(\ref{eq:general}),
$f(z, t) = F(z - st) = F(\xi).$ Substitution into
Eq.~(\ref{eq:general}) with $J = J_{\rm friction}$ yields
\begin{equation}
  \label{eq:steady}
  \left\{A(s) F'(\xi) + B(s) \, {\cal P} \int {F(\xi') \over (\xi -
      \xi')^2 } \, d\xi' + T - \gamma \right\} \Theta[-s F'(\xi)] = 0,
\end{equation}
where ${\cal P}$ denotes the Cauchy principal value and prime denotes
derivative with respect to $\xi.$ The coefficients in
Eq.~(\ref{eq:steady}) are given by
\begin{eqnarray}
  \label{eq:steady_coeffs}
  A(s) &=& \mu - {s^2 \text{sign}(s) (1 - s^2/\kappa)\sqrt{s^2 - 1} -
    \mu s^2 \over 2[1 + (s^2 - 1)(1 - s^2/\kappa)]}\\
  B(s) &=& -{s^2\sqrt{1 - s^2/\kappa}\, [\mu \, \text{sign}(s)
    \sqrt{s^2 - 1} + 1] \over 2\pi[1 + (s^2 - 1)(1 - s^2/\kappa)]}.
\end{eqnarray}
We have shown \cite{ournotes} that a solution to Eq.~(\ref{eq:steady})
exists if and only if there exists $s_{0}$ such that $B(s_0) = 0.$
Steady state solutions can therefore propagate only when $\mu >
\mu_{c} = 1/\sqrt{\kappa - 1}$ and at a velocity $s_0 = -\sqrt{1 +
  1/\mu^2}.$ Furthermore, since $s_0$ is negative, the expression in
the curly brackets in Eq.~(\ref{eq:steady}) must vanish when $F' > 0$.
For loads below threshold, i.e., $T < \gamma$, we find a steady state
solution of the form
\begin{equation}
  \label{eq:kink}
  F'(\xi) = {\gamma - T \over A(s_0)}\, \Theta(\xi).
\end{equation}
This is a starting kink moving in the negative $z$ direction.  A
stopping kink $F'(\xi) = {\gamma - T \over A(s_0)}\, [1 -
\Theta(\xi)],$ and pulses, $F'(\xi) = {\gamma - T \over A(s_0)}\,
[\Theta(\xi)- \Theta(\xi-a)],$ of any width $a$ are also solutions.
These predictions are confirmed by our numerics.

These kink solutions are stable.  When the speed of the backward
moving kink $|s| > |s_0|,$ $B > 0$ and the integral term in
Eq.~(\ref{eq:steady}) diverges logarithmically at the kink tip, $\xi =
0,$ so as to slow it down.  If, on the other hand, $|s| < |s_0|$ the
same term diverges with the opposite sign speeding up the kink.

The motion of a planar crack through a three dimensional elastic solid
is also described by Eq.~(\ref{eq:general}) with $J = J_{\rm
  fracture}(z, t)$ sketched in Fig.~\ref{fig:kernels} (d).  Here,
$f(z, t)$ is the shape of the crack front.  The equation of motion is
the statement that the local elastic energy flux to the crack tip is spent
to create new surfaces there.  The driving force $T$ is related to the
external loads and $\gamma$ is the surface energy \cite{sharad}.  Here
again we find starting and stopping kinks that move with a
characteristic velocity.  Unlike in the case of friction, these kinks
move in either direction along the crack front.

Further progress can be made by realizing that the motion of elastic
boundaries is generically described by Eq.~(\ref{eq:general}).  It
possesses several remarkable features that arise due to the
elastodynamics of the bulk and lead to the existence of steady state
kinks.  First, the equation for the motion of the boundary is first
order in time.  Second, the kernels $J(z, t)$ have the following
properties: (i) $J(z, t < |z|/c) = 0,$ where $c$ is the longitudinal
wave velocity, (ii) the kernel is homogeneous degree $-2,$ so that
$J(z, t) = {j(z/t) \over z^2}$ with $j(0) = 1,$ and (iii) $J$ is
non-monotonic as a function of $t$ at all $z.$ In particular, at a
fixed $z,$ $J$ is negative for some period after the arrival of the
longitudinal sound wave and then changes sign \cite{kernel}.  This
change of sign implies there exists a speed $s_0 < c$ such that for
all $z$
\begin{equation}
  \label{eq:kink_condition}
  \int_{|z|/c}^{|z|/s_0} J(z, t') dt' = 0
\end{equation}
when $J(z, t) = J(-z, t).$ The equation (\ref{eq:kink_condition}) is a
sufficient condition for steady kink solutions that move with speed
$s_0$ in either direction to exist.\footnote{When the kernel is
  asymmetric, kinks may only move in one direction, as in the case of
  friction.}  To see that, we first note that condition (ii) implies
that the steady state equation (\ref{eq:steady}) is generally valid.
Furthermore, we can rigorously show that condition
(\ref{eq:kink_condition}) implies that $B(s_0) = 0.$ Thus steady state
kink solutions exist for all such kernels.

The detailed form of kernels that arise in problems of moving elastic
boundaries depend on various parameters such as the microscopic time
and length scales, and also on reflection of waves from free surfaces,
etc.  We would like to consider a simple model that captures effects
of vectorial elastodynamics to understand the interaction of the kinks
with disorder near the onset of motion.  We therefore study a model
kernel, shown in Fig.~\ref{fig:kernels} (a),
\begin{equation}
  \label{eq:model}
  J_{\rm model} = {1 \over z^2}[-\Theta(t - |z|) + (2 +
  \alpha)\Theta(c_1 t - |z|) - \alpha \Theta(c_2 t - |z|)], 
\end{equation}
with $1 > c_1 > c_2.$ At a fixed position $z$ this model kernel is
zero until time $t = |z|$.  In the time interval $t \in (|z|,
|z|/c_1)$ it is equal to $-1/z^2$.  It then jumps to $(\alpha +
1)/z^2$ until time $t = |z|/c_2$ at which it settles down to its
static value of $1/z^2.$

For $T < \gamma$, the equation of motion (\ref{eq:general}), with the
model kernel, admits starting and stopping kinks that move in both
directions.  Their velocity can be calculated using equation
(\ref{eq:kink_condition}).  It is
\begin{equation}
  \label{eq:s_simp}
  s_0 = \left\{
    \begin{tabular}{lr}
      $\displaystyle{{c_1(\alpha + 1) \over \alpha + 2 - c_1}},$ &
      for $(\alpha + 1)(c_1 - c_2) > c_2 (1 - c_1)$ \\[5pt]
      $\displaystyle{{c_1 \over 2 + \alpha - c_1 - \alpha c_1/c_2}},$
      & otherwise.
    \end{tabular}
  \right.
\end{equation}
This is indeed confirmed by our numerics.

Let us now turn to the effects of disorder which arises due to
spatially varying material properties, such as the coefficient of
friction or the surface energy.  We introduce disorder by making
$\gamma$ in Eq.~(\ref{eq:general}) position dependent.  It is
important to note that in general, the function $\gamma$ depends both
on the coordinate $z$ along the boundary as well as the position of
the boundary $f(z, t)$.  In our simulations, the variable $\gamma(z,
f)$ is uniformly distributed in $[0, \Gamma].$ All lengths are
measured in units of the correlation length of $\gamma$.

We study Eq.~(\ref{eq:general}) numerically by a finite difference
method.  Let us first consider the case the model kernel with $\alpha
= 0.$ We increase the external driving force in small increments from
zero.  These increments are chosen in such a way as to dislodge the
most weekly pinned point of the boundary.  Avalanches begin with two
starting kinks that nucleate there and move outward until they
encounter a tough patch.  Stopping kinks are then nucleated and move
inward until the boundary is arrested stopping the avalanche.  The
dynamic exponent $z$ is defined by the duration $\tau \sim \ell^z$ of
an avalanche of size $\ell.$ The avalanches show a power law
distribution up to a characteristic length $\xi_-$ with longer
avalanches being much rarer.  Below the length scale $\xi_-$ the
boundary is self affine with $\langle |f(z, t) - f(z', t)|^2\rangle
\sim |z - z'|^{2\zeta}$, which defines the roughness exponent,
$\zeta$.  As the load approaches the critical load, $T_{\rm t}^{\rm
  qs}$, $\xi_-$ diverges as $(T_{\rm t}^{\rm qs} - T)^{-\nu}$,
defining the correlation exponent, $\nu.$ Above this load, the
boundary as a whole begins to move with an average velocity, $v \sim
(T - T_{\rm t}^{\rm qs})^\beta$, $\beta$ being the velocity exponent.

\begin{figure}[htbp]
  \begin{center}
    \centerline{\epsfxsize=5in \epsfbox{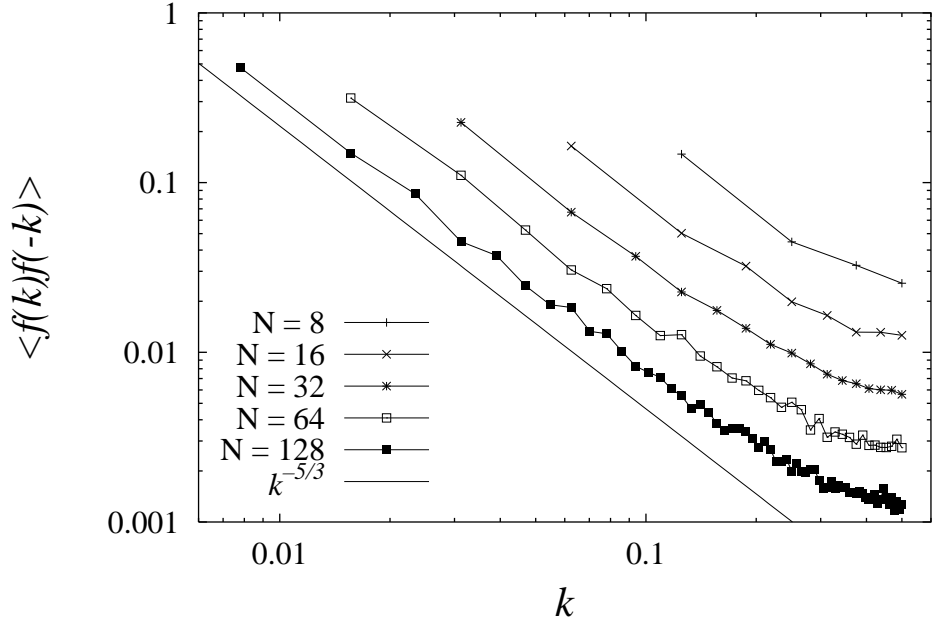}}
    \caption{Boundary statistics $< \! f(k)f(-k) \! >$ just before it
      begins to move for the model kernel with no bump ($\alpha = 0$).
      The solid line is the $k^{-5/3}$ scaling expected from the
      renormalization group arguments.  Length are measured in terms
      of the correlation length of the randomness.}
    \label{fig:ff}
  \end{center}
\end{figure}

\begin{figure}[htbp]
  \begin{center}
    \centerline{\epsfxsize=5in \epsfbox{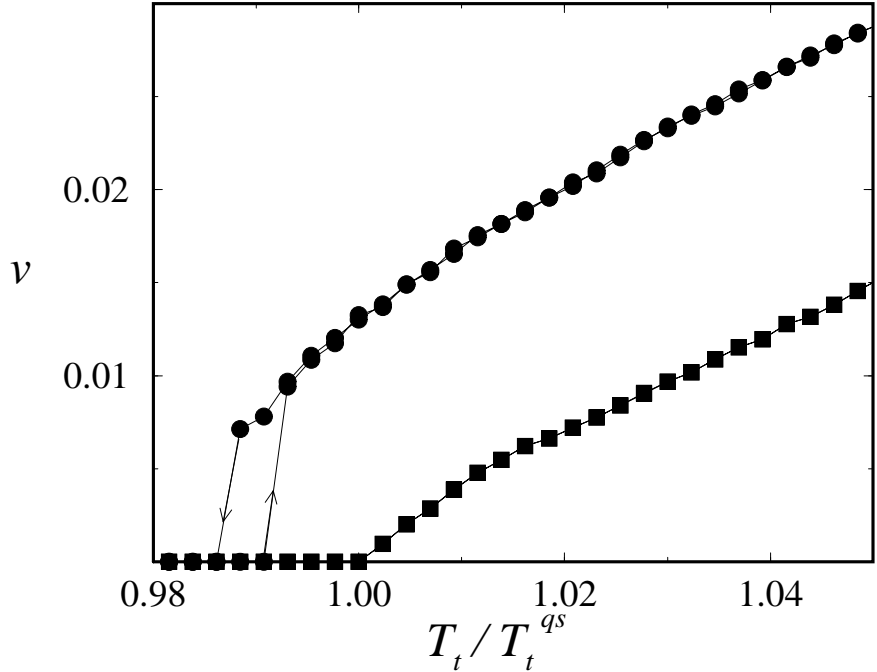}}
    \caption{Average boundary velocity $v$ in units of the fastest
      wave speed as a function of the applied driving $T$ in units of
      the quasistatic threshold load.  The model kernel with $\alpha =
      0$ data is plotted with squares.  When $\alpha > 0$ (circles)
      the velocity is discontinuous and hysteretic.}
    \label{fig:vvsf}
  \end{center}
\end{figure}

The Eq.~(\ref{eq:general}) has been studied extensively for a
quasistatic model where sound waves in the bulk are neglected, so that
$J = J_{\text{qs}}(z, t) = 1/z^2$ \cite{qs}.  Note that
$J_{\text{model}}(\alpha = 0) \leq J_{\text{qs}}$ for all $z$ and $t$,
and for $t > 1/c_2,$ the two are equal.  The no-passing rule
\cite{middleton} implies that the static exponents $\zeta$ and $\nu$
are the same in both cases.  For the quasistatic model, the physics is
determined by a critical fixed point and a $2 - \varepsilon$
renormalization group calculation yields, with $\varepsilon = 1,$
$\zeta = 1/3,$ $\nu = 3/2.$ The dynamic exponents for our model are
different from the quasistatic model and a renormalization group
calculation yields the dynamic exponent, $z = 1$ and $\beta = 1$.
These results agree with our numerics, and the results are shown in
Figs.~\ref{fig:ff} and \ref{fig:vvsf}.

\begin{figure}[htbp]
  \begin{center}
    \centerline{\epsfxsize=5in \epsfbox{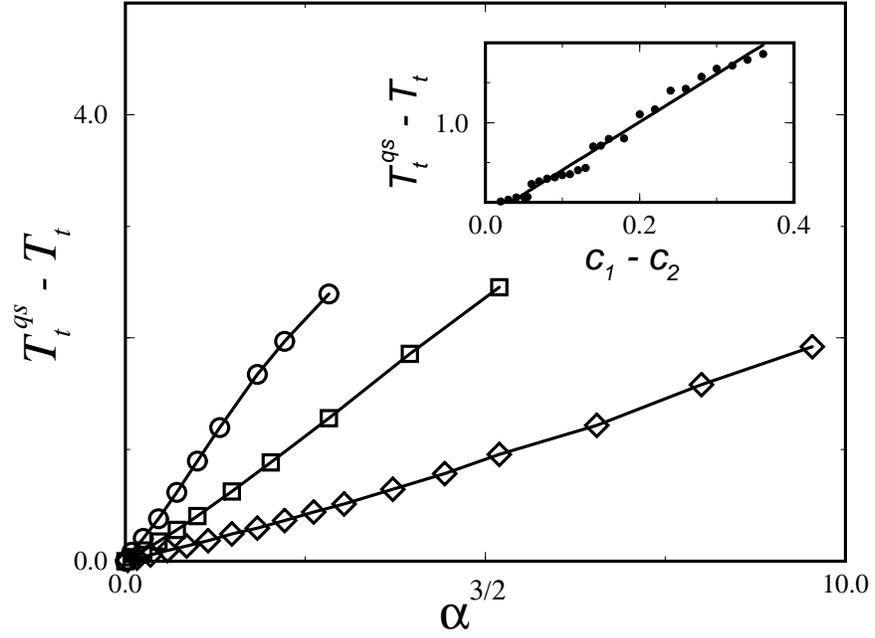}}
    \caption{Threshold force suppression for the model kernel with
      respect to the quasistatic value versus the $3/2$ power of the
      bump height $\alpha$.  The forces $T$ are in natural units.
      Diamonds correspond to $c_2 = 0.417,$ squares to $c_2 = 0.333,$
      and circles to $c_2 = 0.2.$ The inset shows the threshold force
      as function of the $c_1 - c_2.$ In both graphs $c_1 = 0.5$ is
      fixed.}
    \label{fig:scaling}
  \end{center}
\end{figure}

We find that for $\alpha > 0$ the structure of avalanches is the same
as before for small enough loads.  The typical avalanche size
increases with the load.  When it reaches a critical value $\ell_c$
determined by the parameters $\alpha,$ $c_1$ and $c_2,$ the starting
kinks zip through the whole system causing it to move.  We make this
inference from the fact that the threshold load $T_{\rm t}(\alpha,
c_1, c_2),$ is lower than that for the quasistatic kernel $T_{\rm
  t}^{\rm qs}$ and that
\begin{equation}
  \label{eq:scaling}
  T_{\rm t}^{\rm qs} - T_{\rm t}(\alpha, c_1, c_2) \sim \alpha^{3/2}
  (c_1 - c_2) g(c_1),
\end{equation}
as shown in Fig.~(\ref{fig:scaling}).  This scaling form of the
suppression of the threshold load is in agreement with our analytical
results \cite{ournotes}.  Thus, as in the acoustic models in
Refs.~\cite{dsf,dsf1}, we expect a power law distribution of small
events with the same power law as the quasistatic model combined with
characteristic system size events.  Also the velocity jumps to a
finite value at $T = T_{\rm t}(\alpha, c_1, c_2)$ as shown in
Fig.~\ref{fig:vvsf}.  From our study we expect qualitatively similar
behavior near threshold in the case of friction and fracture.

In conclusion, we have demonstrated that vectorial elasticity of the
bulk must be taken into account in order to properly describe dynamics
of elastic boundaries.  Specifically, steadily propagating kinks arise
naturally in elastodynamic models.  Pulse solutions propagating with a
specific velocity, where parts of the boundary are moving can be
constructed from starting and stopping kinks.  Spatial width of the
pulse solutions is arbitrary and must be set by some nonlinearity in
the system or the reflection of waves from free boundaries.  These
pulse solutions should be observable in experiments on fracture
similar to those of Schmittbuhl \cite{schmittbuhl}.  These kinks can
presumably be triggered at an edge of the sample.  Simulations of
solid on solid friction including the bulk elastodynamics
\cite{stefanprivate} suggest that the key features of the kernels
obtained from a continuum treatment persist and hence kinks solutions
must still exist though the velocity of motion of these kinks will be
determined by the detailed shape of the kernel.  Preliminary results
indicate that the effects of sound damping in the bulk is to set a
spatial extent for the pulses, but the details remain to be explored.
More complicated friction laws, such as different static and dynamic
friction coefficients, lead to complex behavior left for future
studies.  However, we expect kink solutions and the qualitative
threshold behavior to persist.

In systems with disorder, the dynamics near the onset of motion is
controlled by nucleation of starting kinks by weak patches and of
stopping kinks by tougher regions.  Once an avalanche reaches a
critical size which is set by the details of the kernel, it runs away
spanning the entire system.  Therefore, there should be an excess of
system size avalanches beyond the crossover scale.  Even when one
includes the finite response time of a point on the boundary to
changes in the stress, the same physics persists.  Since avalanches
beyond a critical size run away, rare weak patches should play an
important role that has yet to be explored.

In this letter we restricted our attention to one dimensional
boundaries.  When the spatial dimensionality of the boundary is two,
"kink-on-kink" solutions arise.  Thus, a possible mechanism for
stopping a spreading circular earthquake, is a stopping kink-on-kink
that propagates around the earthquake front.  Questions about the
dynamics of these kinks as well as their interaction with disorder
remain open.  However, models in which kinks are featured as the
fundamental excitations offer significant advantages over full scale
three dimensional numerical simulations and should be studied.

We would like to thank Daniel S. Fisher, Jim Rice, Jean Carlson and
Jim Langer for extensive discussions.  This work was supported by
National Science Foundation via PHY94-07194 and DMR-9510394.  AL
wishes to acknowledge support of the Department of Energy via grant
No. DE-FG03-84ER45108.

\bibliographystyle{prsty}
\bibliography{kinks}

\begin{thebibliography}{10}
\bibitem{other_laws}
If the coefficient of friction $\mu$ is velocity or slip dependent, the only
  change to first order in the speed of the boundary, is in the coefficient of
  the left hand side of Eq.~(\ref{eq:general}).
\bibitem{footnote2}
In fact, when $T > \gamma$, there is a propagating mode in the velocity
  fluctuations of the boundary for $\mu$ greater than a critical value. This
  mode can be excited by any heterogeneity in the friction constant or spatial
  variations in the driving load. Thus the physics even above the threshold
  load $\gamma$ should be interesting.
\bibitem{kernel}
When the symmetry of the problem is such that one component of the displacement
  field remains decoupled from the others, acoustic like kernels with $J(z, t)
  \sim {t \over z^2} {1 \over (t^2 - z^2)^{1/2}}$ arise for the equations of
  motion of these components, as that for a crack front under tearing loading
  or the component of the displacement transverse to the direction of the shear
  in the case of friction. There are two necessary conditions for
  elastodynamics to give rise to kernels with property (iii). First, the
  co-dimension of the elastic interface has to be spanned by components of the
  displacement field. And second, the displacement field components along the
  boundary must not be decoupled from the bulk.
\end{thebibliography}

\end{document}